\documentclass{cimento}
\usepackage{graphicx}  
\def\swift{{\em Swift\/}}
\def\sax{{\em BeppoSAX\/}}
\def\gro{{\em Compton Gamma Ray Observatory\/}}

\def\hete{{\em HETE 2\/}}
\def\fermi{{\em Fermi\/}}
\def\agile{{\em AGILE\/}}

\title{Gamma Ray Burst origin and their afterglow: story of a discovery and more}
\author{Enrico~Costa\from{ins:1}
        \atque
Filippo~Frontera\from{ins:2}
}
\instlist{\inst{ins:1} INAF, IASF, Rome, Italy
  \inst{ins:2} University of Ferrara, Physics Department, Via Saragat, 1, 44100 Ferrara, Italy
}
\PACSes{\PACSit{95.85.Nv}
\PACSit{95.85.Pw}
\PACSit{98.70.Rz}
\PACSit{98.80.Es}
}
\begin{document}

\maketitle

\begin{abstract}
In this paper we review the story of the \sax\ discovery of the Gamma Ray Burst afterglow and their cosmological
distance, starting from their first detection
with Vela satellites and from the efforts done before \sax. We also discuss the consequences of the \sax\ discovery, 
the issues left open by \sax, the progress done up to now and its perspectives. 
\end{abstract}

\section{Introduction}

Gamma Ray Bursts (GRBs) are the most fascinating events in the Universe. They are sudden bright flashes 
of gamma--ray radiation of celestial origin, with variable
duration from milliseconds to several hundreds of seconds. In rare cases time duration up to thousands of seconds
has been observed. Most of the emission is made of photons with energies from several keV to tens of MeV 
and above.
With the current instrumentation their observed occurrence rate is 2--3 per day over the entire sky.
Their arrival time is impredictable as it is impredictable their arrival direction.
When they are on, their brightness overcomes any other celestial gamma-ray source.
They were discovered by chance at the end the 60's with the American Vela spy satellites, 
devoted  to monitor the compliance of the Soviet Union and other nuclear-capable states with the 1963 
"Partial Test Ban Treaty".
 The American militaries kept this discovery in their drawers until they decided to talk 
about that with scientists colleagues in Los Alamos laboratories, who published their discovery
in 1973 \cite{Klebesadel73} with the title "Observations of Gamma-Ray Bursts of Cosmic Origin".
Once discovered, the first questions were: which are their progenitors? Are they normal stars or compact stars,
like white dwarfs (WD), neutron stars (NS) or Black Holes (BH)? or more simply they are something that occurs in the
interstellar medium? or in comets? At which distance are their sites? Are local sites? Are their sites in our Galaxy 
or in extragalactic objects? Which is the power involved?

The solution of these issues implied  complex observational problems, like the accurate 
localization of the events. This was  a tough task in the gamma--ray energy band, where detectors could
give only coarse source directions. A possible solution of this problem could be the search of GRB 
counterparts at longer wavelengths, that could be a new born source or an already existing one.
 
Many satellite missions (e.g., the Russian satellites Venera 11, Venera 12, Prognoz 6, Prognoz 9, Konus, Granat, 
and the American Pioneer--Venus Orbiter and Solar Maximum Mission) 
which included instrumentation devoted to the detection of GRBs (see, e.g., Refs.~\cite{Niel76,Barat81}) 
were performed, but a very small progress in the
GRB origin was obtained (see, e.g., Ref.~\cite{Epstein88} and the very complete review by 
Vedrenne and Atteia \cite{Vedrenne09}). 
Both their spatial distribution and their distance
was a persistent  mystery. Their origin was reminiscent of the great debate in 1920s about the local or 
extragalactic origin of spiral nebulae. 
The localizations were very coarse and no counterpart at longer wavelengths was found (see, e.g., Ref.~\cite{Hudec87}). 
The most accurate localizations, with error boxes even of a few square arcminutes, were obtained using
the triangulation method. This consists in the accurate timing of GRBs with omnidirectional detectors aboard at least
three satellites. From the time delay in the arrival time of the same event by these satellites it is possible to derive
the arrival direction of the event. However these localizations were obtained after long times, from months to 
years, from the GRB occurrence (e.g., Ref.~\cite{Atteia87}). 

Many theoretical models on  GRB progenitors were worked out, with the largest consensus being obtained by the models 
which assumed that GRBs originate in Galactic disk neutron stars.

The largest observational effort  was performed with the BATSE experiment  aboard the NASA \gro\ satellite (CGRO) 
\cite{Fishman94}, with a very wide field of view (2$\pi$) (see below). In spite of the great wealth of information, 
the sites of GRBs were not discovered.
The real revolution came with the \sax\ satellite that, in only 6 months from its launch (30 April 1996), allowed
to discover the X--ray counterparts of GRBs and their distance. After about 30 years from their discovery, we
eventually learnt that GRBs are huge explosions in galaxies at cosmological distances.

In this paper we review the major results obtained in the pre--\sax\ era, the \sax\ revolution and its major
results obtained. We will also discuss the consequences of the \sax\ discovery, the issues left open  and the
further  progress obtained with later missions devoted to GRBs. Finally we will discuss some relevant still open issues and 
the prospects to solve them.

\section{The pre--\sax\ era}

The long epoch of missions centered on the measurement of GRBs in the hard X--/soft gamma--ray band was 
completed by the BATSE experiment onboard CGRO. 

It consisted of eight couple of flat detectors completely open, located on 
all corners of the CGRO spacecraft, which had the shape of an octahedron. Each couple was made of NaI(Tl) scintillators: 
a Large Area Detector (LAD) and a Spectroscopy Detector (SD) \cite{Fishman94,Preece00}.
Each LAD  had a large collection area (2000 cm$^2$) and a small thickness (1.27 cm), while each SD had a small 
area (127~cm$^2$) and a large thickness (7.2 cm). BATSE had an unprecedented sensitivity and 
the most achievable uniform response to all sky directions. It was particularly suited to perform systematic studies
of GRBs with a good control of the angular coverage. The nominal energy passband was 25 to 1800 keV. The LAD spectra 
were singly accumulated into 128 quasi-logarithmic  energy channels, while the SD spectra were singly accumulated 
into 256 quasi-linear spectra. Exploiting their location and the angular response of the detectors (approximately a 
cosine function at lower energies), the detected GRBs could be localized on the sky with an accuracy 
of about 2 degrees  at best for the brightest events \cite{Briggs99}.
BATSE provided the best and most complete set of data on GRBs before the \sax\ launch.

Launched in April 1991 CGRO operated in low Earth orbit (LEO)
for over 9 years until it was de-orbited in June 2000. In these 9 years, BATSE detected 
2702 GRBs \cite{Batse_cat}. Thanks to BATSE, several  properties of the GRB prompt emission
were observed and investigated (see, e.g., Ref.~\cite{Paciesas04}, for a summary).

\subsection{Main spectral and temporal properties derived with BATSE}
 
The main GRB properties discovered with BATSE are summarized below.

\subsubsection{Bimodality of GRB durations}

\begin{figure}[!t]
	\begin{center}
	\includegraphics[width=0.6\textwidth]{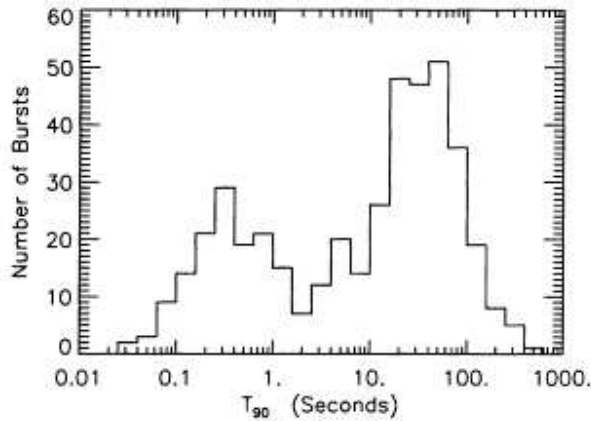}
		\caption{Distribution of $T_{90}$ for 427 GRBs detected with BATSE \cite{Meegan96}. $T_{90}$ is defined
as the time during which the GRB cumulative counts increase from 5\% to 95\% of the total detected counts.}

	\label{f:t90_distr}
	\end{center}
\end{figure}

It was found with BATSE (see, e.g., Ref.~\cite{Meegan96}) and later confirmed by \sax\ \cite{Frontera09}, 
that GRB durations  show a bimodal distribution, with two broad peaks centred at about 0.3 s and 20 s with a minimum 
at about 2 s. This bimodal distribution (see Fig.~\ref{f:t90_distr}) separates GRBs in two broad categories, short bursts 
($<$2~s) and long bursts ($>$2~s).  

%
%

\subsubsection{GRB fine temporal stucture}

\begin{figure}[!t]
	\begin{center}
	\includegraphics[width=0.6\textwidth]{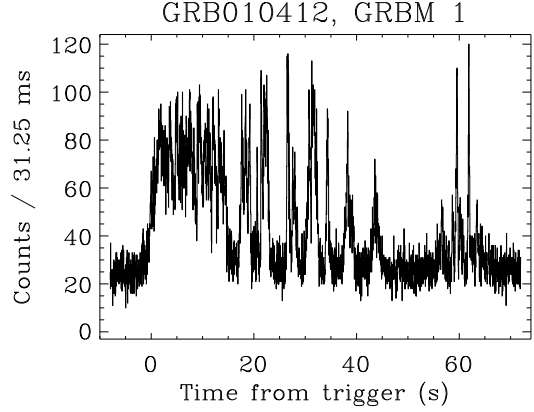}
		\caption{An example of a GRB  occurred on 2001 April 12 (named GRB 010412) with numerous spike structures.
		It was detected with the \sax\ GRBM \cite{Frontera09}.}
	\label{f:time_var}
	\end{center}
\end{figure}

Thanks to the large effective area of BATSE, it was soon found (Ref.~\cite{Fishman92,Walker00}) that some GRBs 
show spiky structures with durations down to ms time scales. A pecular example of a GRB with numerous spiky structures
is shown in Fig.~\ref{f:time_var}. From these results,
it was inferred a small size scale of the source ($R \sim c\Delta t \sim 10^7$~cm).

%

\subsubsection{Spectral distribution of the radiation}

\begin{figure}[!t]
	\begin{center}
	\includegraphics[width=0.6\textwidth]{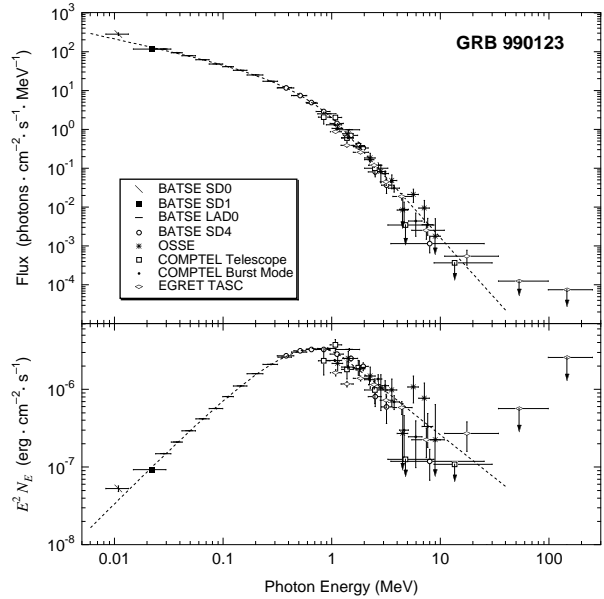}
		\caption{An example of GRB photon spectrum and the corresponding $E^2 N(E)$ function.  It
	shows the spectrum of the GRB occurred on 1999 January 23 (GRB\,990123), that was prompty detected and
 localized with \sax. Reprinted from Ref.~\cite{Briggs99}.}
	\label{f:nuFnu}
	\end{center}
\end{figure}

Using BATSE results, Band et al.\cite{Band93} found that the best description of the time averaged 
photon spectrum of a GRB, from 20 keV up to a few MeV, was a smoothly broken power--law, 
now dubbed {\bf Band function} (see top panel of Fig.~\ref{f:nuFnu}):
\begin{eqnarray}
N(E) = A 
\cases{\displaystyle 
\left(\frac{E}{100keV}\right)^\alpha \exp{\left({-E/E_{0}}\right)},
\cr
\cr
{\rm for}~(\alpha - \beta)\cdot E_{0} \hbox{  }\ge\hbox{  } E;
\cr
{\displaystyle
\left[\frac{(\alpha - \beta) E_{0}}{100keV}\right]^{\alpha - \beta} \exp{(\beta - \alpha)}
\cdot \left(\frac{E}{100keV}\right)^\beta} 
\cr
\cr
~~{\rm for}~(\beta - \alpha) \cdot E_{0}\hbox{  }\le \hbox{  } E},
\end{eqnarray}

where $\alpha$ and $\beta$ are the power--law
low energy (below E$_0$) and high energy (above E$_0$) photon indices, 
respectively, and A is the normalization parameter.
The values of these parameters change from a GRB to another, with typical values
of $\alpha = -1$, $\beta = -2.3$ and $E_0 = 150$~keV.
If this is the spectral description, the spectra  are clearly nonthermal. This result
is now widely confirmed (see, e.g., Ref.~\cite{Guidorzi11}).

The non thermal shape can be better seen by plotting $E^2 N(E)$ vs. E, that gives the emitted power per energy decade.
The new function, also dubbed $\nu F_{\nu}$ spectrum (where $F_{\nu}$ is the energy spectrum), generally shows a 
maximum at photon energy $E_p = E_0 (2 + \alpha)$ if $\beta < -2$. This is what generally happens for 
GRBs (see bottom panel of Fig.~\ref{f:nuFnu}).

%

\subsection{Angular and intensity distribution of GRBs}

The most crucial results obtained with BATSE were actually two: the isotropic distribution of GRBs in the sky and their
non homogeneous distribution in distance.

\subsubsection{Angular distribution of GRBs}

\begin{figure}[!t]
	\begin{center}
	\includegraphics[width=0.6\textwidth]{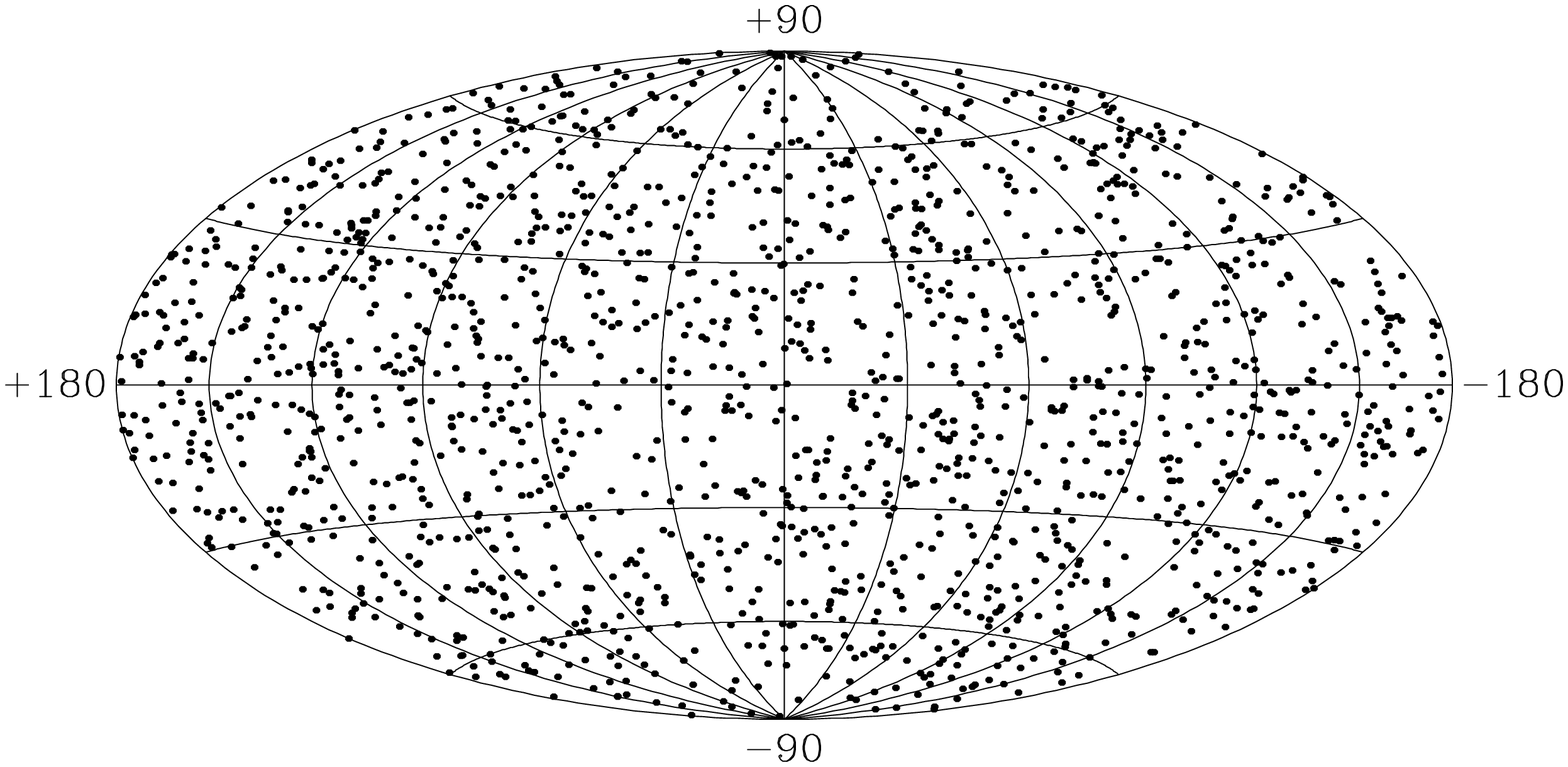}
		\caption{Sky distribution of the 1637 GRBs detected with BATSE from April 1991 to August 1996,  
	in an Aitoff--Hammer projection in Galactic coordinates. Reprinted from Ref.~\cite{Paciesas99}.}
	\label{f:sky_distr}
	\end{center}
\end{figure}

In spite that BATSE did not have the capability of accurately determining the arrival direction of GRBs (see above),
it settled a long awaited issue: the angular distribution of GRBs in the sky. The result was that GRBs 
are isotropically distributed in the sky. Figure~\ref{f:sky_distr} shows the distribution obtained with BATSE on 
the basis of 1637 events. From this result, the assumption that GRBs could have origin in the disk of our Galaxy was 
definitely rejected. 
All GRB models that involved  Galactic disk source populations disappeared.

%

\subsubsection{Intensity distribution of GRBs}

\begin{figure}[!t]
	\begin{center}
	\includegraphics[width=0.6\textwidth]{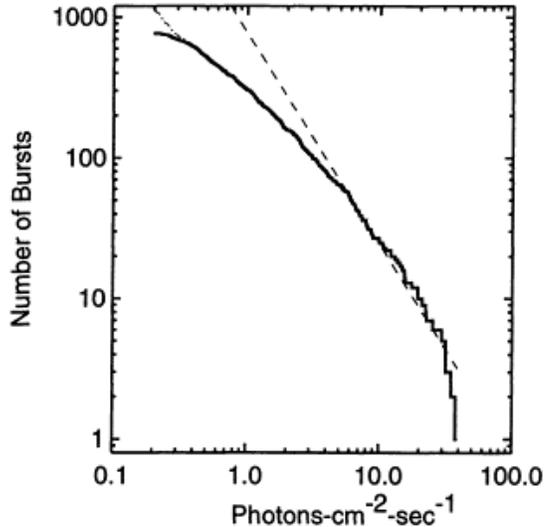}
		\caption{Integral $\log N$--$\log P$ distribution of 772 GRB detected with BATSE in the range 50--300 keV band.
The peak flux $P$ is integrated over 1.024 s. The dot-dashed line gives the correction for instrumental trigger efficiency,
while the dashed line gives the power--law slope ($-3/2$) expected in the case of an homogeneous distribution of GRBs in space.
Reprinted from Ref.~\cite{Meegan96}.}
	\label{f:logN-logP}
	\end{center}
\end{figure}

Thanks to its high sensitivity, BATSE was capable to provide another key result: the paucity of weak bursts. 
This effect is apparent in the distribution of the number of GRBs as a function of their flux.
Given the difficulty of deriving an unbiased estimate of the GRB flux (the GRB tail estimate is a fraction of the
background level), the GRB peak flux $P$ was adopted. An example of this distribution is shown in Fig.~\ref{f:logN-logP}.
In an Eucledian space, if the burst sources are uniformly distributed in distance and have either the same luminosity 
(standard candles) or have a non evolving luminosity function, 
the number of visible sources would increase as the cube of their distance ($N \propto d^3$), while the source flux 
would decrease as the square of their distance 
($F \propto d^{-2}$). As a consequence the number of burst sources with flux greater than a given value $F$
would be given by
\begin{equation}
N(>F) \propto F^{-3/2}
\label{e:logN-logF}
\end{equation}

Thus, if there was a deviation from Eq.~\ref{e:logN-logF}, it would mean a non--homogeneous distribution of
the sources with distance.

Actually a deviation from Eq.~\ref{e:logN-logF} could also be consistent with an origin of GRBs in the 
disk of our Galaxy (the first assumed GRB site),
if it could not be demonstrated  that the weak bursts have an isotropical distribution in the sky. Indeed, the
paucity of weak bursts, found even before the BATSE era (see, e.g, Ref.~\cite{Mazets85}) was considered as
a demonstration that GRBs had origin in the disk of our Galaxy.
The great result obtained with BATSE was that not only all GRBs, independently of their brightness, but also weak bursts,
 were isotropically distributed in the sky (see Fig.~\ref{f:sky_distr}). 
 
%

With these results, GRBs could be  local, could have origin in an extended halo of our Galaxy with a typical source 
distance of about 100 kpc (see, e.g., \cite{Hartmann94}), or could be extragalactic 
(see, e.g., Ref~\cite{Paczynski86}). They could also have origin in a combination of these sites. No decisive result was 
obtained that could solve the mystery of the GRB sites. Also the theoretical astrophysical Community 
was divided on the GRB site issue: an extragalactic origin was strongly 
defended by, e.g., Paczynski \cite{Paczynski95}, while a Galactic origin was equally defended by, e.g., 
Lamb \cite{Lamb95}. In 1995 Bohdan Paczynski wrote \cite{Paczynski95}: "At this time, the cosmological distance 
scale is strongly favored over the Galactic one, but is not proven. A definite proof (or dis-proof) could 
be provided with the results of a search for very weak bursts in the Andromeda galaxy (M31) with an 
instrument ~10 times more sensitive than BATSE. If the bursters are indeed at cosmological distances then 
they are the most luminous sources of electromagnetic radiation known in the Universe. At this time we have no clue as 
to their nature, even though well over a hundred suggestions were published in the scientific journals. 
An experiment providing ~1 arc second positions would greatly improve the likelihood that counterparts of 
gamma-ray bursters are finally found. A new interplanetary network would offer the best opportunity."
In the same year Don Lamb wrote \cite{Lamb95}: "We do not know the distance scale to gamma-ray bursts. Here I discuss 
several observational results and theoretical calculations which provide evidence about the distance scale. 
First, I describe the recent discovery that many neutron stars have high enough velocities to escape from 
the Milky Way. These high velocity neutron stars form a distant, previously unknown Galactic "corona." 
This distant corona is isotropic when viewed from Earth, and consequently, the population of neutron stars 
in it can easily explain the angular and brightness distributions of the BATSE bursts."

The solution of the mystery about the GRB sites started with the discovery of an X--ray and an optical counterpart 
of the GRB event occurred on 28 February 1997 thanks to the \sax\ satellite.

\section{The \sax\ era}
\label{s:sax}

\subsection{Initial \sax\ motivations}

SAX (Satellite per Astronomia X) was  proposed in 1981 to the
National Space Plan of CNR, later evolved into the Italian Space
Agency(ASI). The main goals were:
\begin{itemize}
\item Study  of celestial X-ray sources in a broad energy band (0.1-300
keV) with narrow field instruments; 
\item X-ray (2-30 keV)
monitoring of the sky, in particular of the Galactic plane.
\end{itemize}

When SAX was proposed, X-ray Astronomy was a 20 year aged science.
The satellite-borne collimated proportional counters were sensitive in 
the band 2-10 keV, while the balloon borne instrumentation was providing most of 
information in the 30-200 keV band. The later introduction of grazing incidence optics 
in satellite missions like {\em Einstein} \cite{Giacconi79} substantially
increased the sensitivity to detect, locate and resolve X-ray sources, but
also resulted into a shift of the operational energy band toward
lower energies ($<$3~keV). With the sensitivity growing, also spectra appeared more
complex. Moreover all the already flown X--ray missions detected a celestial
source variability at all time-scales and at all energies. As variability and  broad band  
were not covered by telescopes, the main target of SAX, later dedicated to Giuseppe
(Beppo) Occhialini, was to cover in a narrow field an unprecedented energy
range with a sensitivity as balanced as possible 
and to contemporarily monitor wide portions of the sky for variability studies. 

%
\begin{figure}[!t]
    \begin{center}
    \includegraphics[width=0.6\textwidth, angle=0]{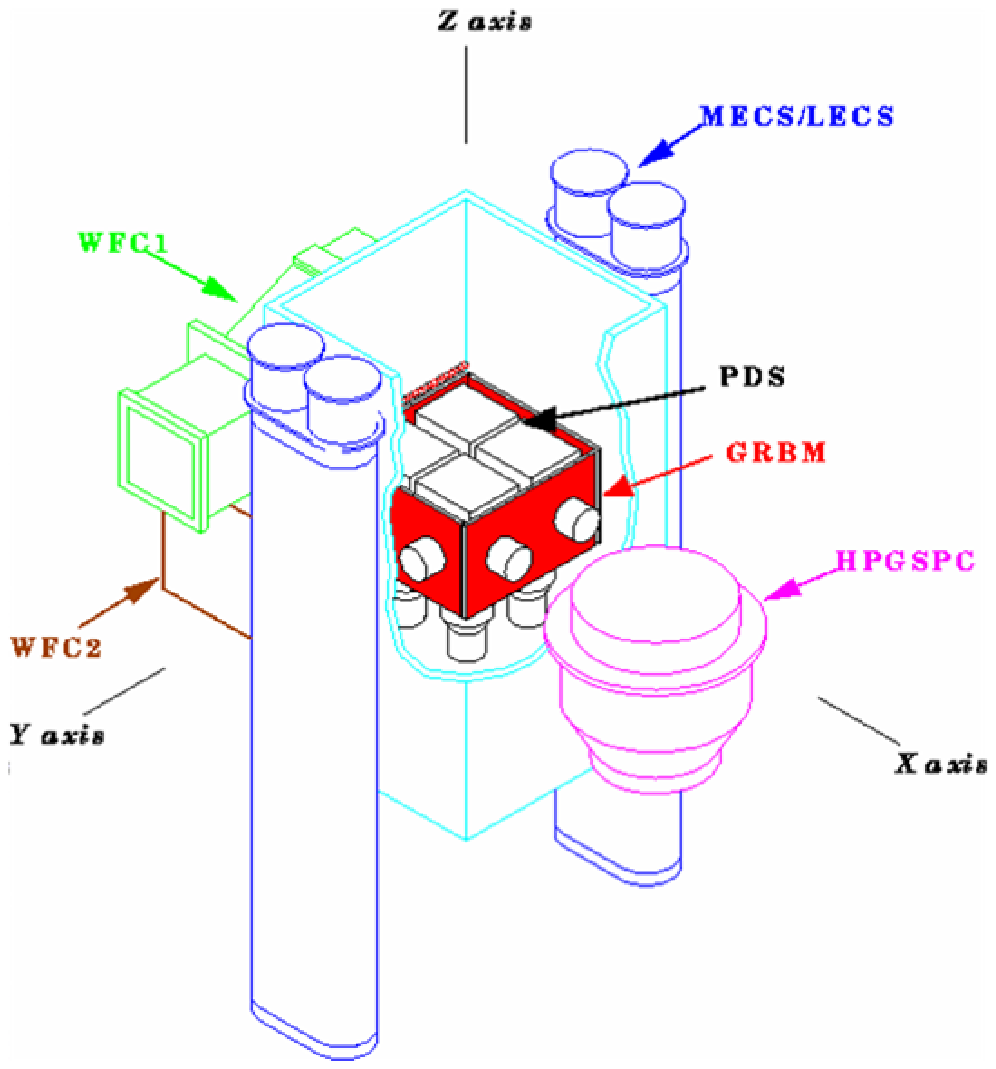}
        \caption{The \sax\ payload.}
    \label{f:SAXpayload}
    \end{center}
\end{figure}

\subsection{The Scientific Instruments of SAX}

To achieve the mentioned goals, the \sax\ payload (see Fig.~\ref{f:SAXpayload})  included
Narrow Field Instruments (NFIs) and Wide Field Cameras (WFCs) \cite{Boella97b}.

The NFIs were:
\begin{itemize}
 \item 4 focusing telescopes LECS+MECS (0.2-10 keV), PI G. Boella
 \cite{Boella97a, Parmar97};
 \item HPGSPC (4-60 keV), PI G. Manzo \cite{Manzo97};
 \item PDS (15-200 keV),   PI F. Frontera (Deputies: E. Costa, D. Dal Fiume) \cite{Frontera97}.
\end{itemize}

The WFCs (2-28 keV, PI R. Jager) consisted in two position
sensitive proportional counters coupled with coded
aperture masks,  capable to provide  an angular resolution of 3--4 arcmin in a field of
view of 20$^{\circ}\times 20^{\circ}$ (fwhm) \cite{Jager97}.

The rationale to have aboard the same mission two WFCs, pointing towards
two opposite directions perpendicular to NFIs, was to monitor the
status of several variable sources and to have the possibility to
perform shorter pointed observations at particular states of them.
A case of particular interest would be that of new transient
sources. 

Notwithstanding this approach, the original proposal did
not include GRBs as science goal. This was due to the fact that in
the early '80s our knowledge about GRBs was strictly confined to
gamma-rays or very hard X-rays. It was not clear whether an X-ray
phenomenology should be expected. For sure, GRBs had no outstanding
phenomenology in X-rays, and no transient phenomenon had been
associated to a GRB until that time. Moreover, due to the wide
spread of interpreting models it was not sure whether any delayed
emission would be detected.

\subsection{The addition of the Gamma Ray Burst Monitor}

In 1984, two years after the SAX approval, during the so
called  phase A study, the Principal Investigator (PI) of the PDS, F. Frontera,
proposed to add to the SAX instrumentation an additional equipment
that could convert the four detectors used as active
anti-coincidence shields surrounding the PDS into a Gamma Ray Burst
Monitor (GRBM). 
The fact that the axis of two of these shields was parallel to that of the two WFCs, 
was indeed very suggestive: some GRBs (the estimate was 2-3 per year) 
could enter into the common field of view of WFC and GRBM and thus they could be 
identified by GRBM and accurately positioned (within 3-5 arcmin) by WFCs. Indeed, 
while in the energy band of GRBM, GRBs could be univocally identified, in that of  WFCs
they could be confused with other transient phenomena. Obviously it was required  to develop, 
among other things, an in-flight trigger system and a proper electronic chain.

In 1990 the anticoincidence detectors with dedicated electronics
was approved and identified as a further instrument of \sax, the
GRBM, and eventually adopted in the
baseline configuration. 
In the further evolution of the project
the design of the GRBM was better defined and improved, in the
context of limited resources, and without interfering with the
function of anti-coincidence that was always the primary driver of
the detector design. A GRBM description can be found in Ref~\cite{Costa98}.

Synthetically, the GRBM included:
\begin{itemize}
 \item A trigger system to identify GRB events
 \item 4 electronic chains for getting continuously spectra and ratemeters
from each GRBM unit;
 \item Gain monitoring;
 \item In case of a trigger, high time resolution ratemeters (down to 0.5 ms).
\end{itemize}

The trigger condition, for each detector, was based on a rapid
change in the rate with respect to a moving average of previous counts.
The trigger coincidence from at least two detectors was requested to discriminate
GRBs from magnetospheric events and from phosphorescence activated by
high atomic number particles of cosmic rays crossing the shields (see, e.g., \cite{Frontera81}).
The latter are the most elusive and can mimic
a short GRB. In fact, whenever the particle crosses two shields, two
coincident signals lasting few milliseconds are generated and the
trigger condition, as defined by the logics of the on board
electronics, is fulfilled. 
GRBM prototype was developed in our Institutes, also with repeated
testing with the SATURNE proton accelerator (Saclay). The Flight
model was developed by Laben Company under the supervision of the
PDS PI.

Thanks to different orientations of the GRBM units, the different exposed 
area to the same GRB  could be exploited to obtain the GRB direction
with an accuracy (few degrees) sufficient for deriving photon spectra
\cite{Pamini90}. GRBM detectors were located in the center of the satellite and the
implementation of this function required a very detailed
description of the whole SAX satellite based on both simulations
and calibrations.

A calibration campaign was performed at ESTEC (Noordwijk, NL)
after the integration of the instrument in the satellite\cite{Amati98}. In parallel
a very extended and detailed Monte Carlo code was built.

\subsection{BeppoSAX in Orbit}
SAX was launched on 30 April 1996 from Cape Canaveral with an
Atlas-Centaure rocket. After a commissioning phase lasted 2 months
and a Science Verification Phase (SVP), lasted 3 months, the operational
phase started on October 1996. 

During SVP, all GRBM programmable parameters, such as thresholds and trigger 
conditions, were set, and a certain number of GRBs was detected. This instrument set
up, that was in practice kept unchanged during the following
five years, resulted in a good confidence on how to discriminate
GRBs from magnetospheric events and phosphorescence. 
A rate of spurious events (about one per orbit) was still there.
This rate was not negligible, when compared with the trigger rate from
\textit{real} bursts, that was about one each three days. 
In order to select the true triggers, duty scientists of the Science 
Operation Center (SOC) would first decide that the burst was likely real, then
they would analyze the light curve of both WFCs, searching for
an excess of counting rate in correspondence with the GRB. If found,
an image of detector at the time of the excess would be accumulated 
and deconvolved to produce an image of the sky. If the excess was associated to a
flaring source, a point-like source would show-up in this sky
image. It must be clear that the \textit{blind search} for the
burst in the WFCs was out of question because the process of accumulation
and deconvolution of the image required tens of minutes. Moreover
we know, a posteriori, that without an external temporal
indication like that given by  GRBM, the discovery of a GRB
from the image, especially at the edge of the field of view, would be
difficult.

\subsection{The first GRB and the set up of a prompt follow--up procedure}

The first GRB in the field of view of WFC was detected on July
20, 1996, but it could be accurately localized only 20 days after
the event. About one month from the burst, the
NFIs were pointed to the GRB direction but no X-ray counterpart was
found \cite{Piro98a}.
 
From this experience it was clear that a possible residual X--ray radiation, 
if any, could have been found only in the case it was possible to promptly point 
the NFIs along the direction of well localized events.
From the analysis of the needed operations to perform a prompt follow--up,
it was clear that the time needed could be significantly shortened. 
Actually, in the case of BeppoSAX, the various teams involved in a fast follow-up (SOC,
Operation Control Center, Scientific Data Center) were in tight connection 
and physically in the same location. In the same place a strong team of instrument experts
could converge. By analyzing the needed operations and
short-cutting all  the interfaces, it was possible to define a
procedure that would start from the arrival of data to the SOC. The
steps were:
\begin{itemize}
 \item Display of GRBM light curves around the trigger time to
discard fake events; 
\item In the case of a GRB detection by GRBM, display of WFC1 and WFC2 data to
search for an X-ray excess  at the same time;
 \item In the case an X--ray excess was found, data accumulation of the interested WFC at the burst time and
deconvolution of the detector image to produce a sky image. Search
for a point-like source in this image. 
\item In the case a source image was found, data accumulation and
deconvolution before the trigger and after the trigger to verify
that the point source was a transient at the time of the burst.
 \item From the deconvolved map and satellite attitude, determination of 
the coordinates of the transient source in the sky reference frame.
\end{itemize}
A re-pointing of BeppoSAX was immediately planned and actuated. 
At the very beginning of the mission, at 90$\%$ confidence level the WFC bursts could be 
localized within a circle of 5 arcmin radius. 
Given that the field of view of the MECS (2--10 keV) was of 50 arcminutes 
diameter, a source localized by the WFC would fall in any case within
the field of view of this telescope.

\subsection{The first afterglow}

The first time this procedure was applied was January 11, 1997. A
bright burst detected by the GRBM was localized in one of the WFCs.
Following the procedure, the NFI telescopes were pointing to the GRB direction 16 hours
after the burst. No source associated
to the burst was found \cite{Feroci98}, but this first experience
helped us to further improve the procedure. Also the localization capability 
of the WFCs shrunk to 3 arcmin radius.

\begin{figure}[!t]
    \begin{center}
    \includegraphics[width=0.6\textwidth]{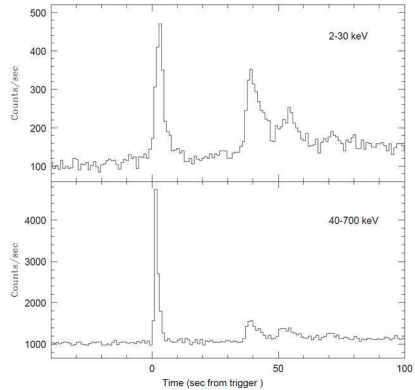}
        \caption{Light curve of GRB\,970228 as seen by GRBM (40--700 keV) and 
	WFC (2--30 keV). Reprinted from Ref~\cite{Costa97}.}
    \label{f:970228_lc}
    \end{center}
\end{figure}

The second burst was detected on 1997 february 28 \cite{Costa97a}.
It consisted of one bright peak, trailed by a train of three more
peaks of decreasing intensity (see Fig.~\ref{f:970228_lc}).

%

\begin{figure}[!t]
    \begin{center}
    \includegraphics[width=0.6\textwidth, angle=270]{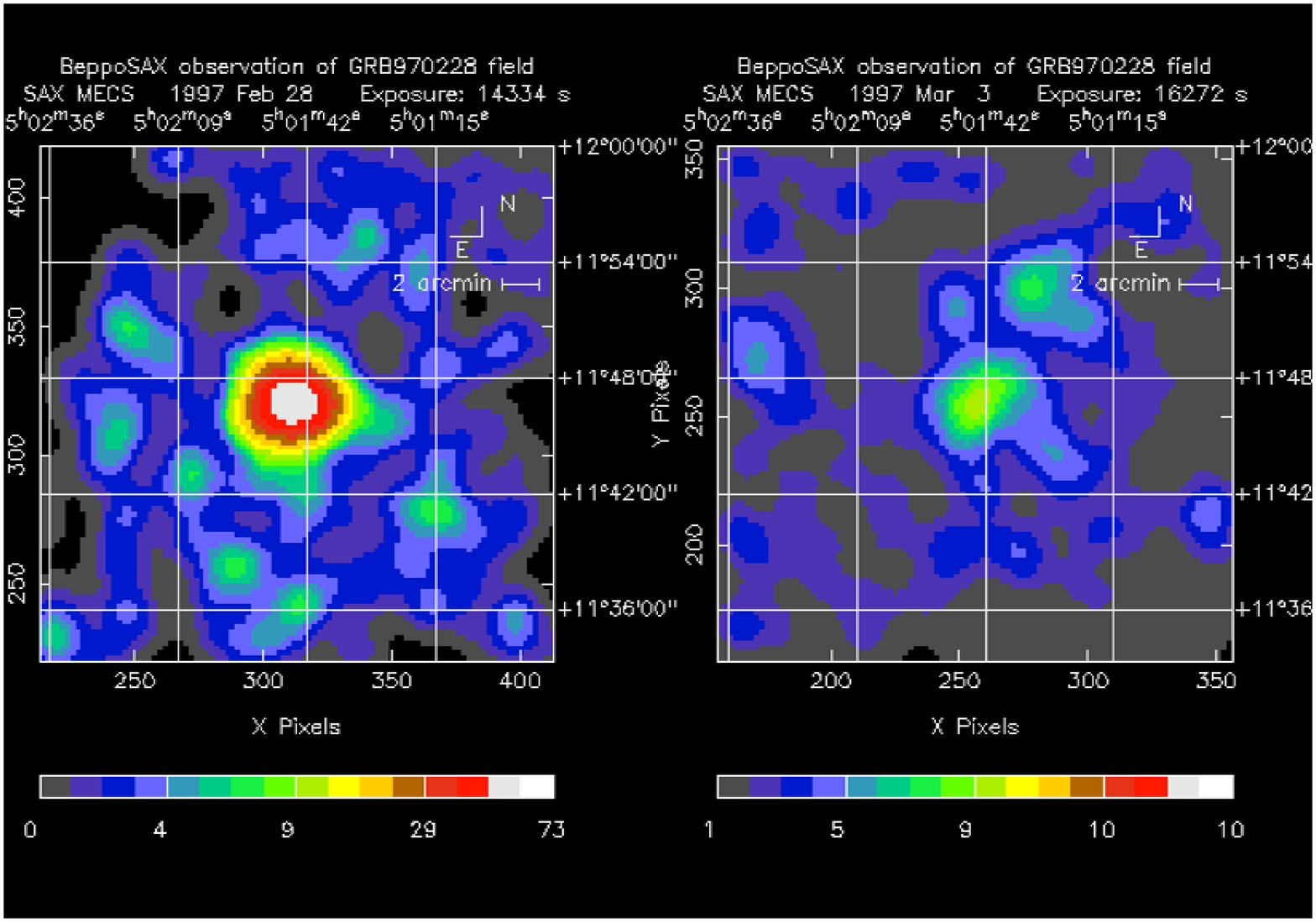}
        \caption{The images of GRB970228 detected with MECS at
        two different epochs. The image on the left includes data
        from 8 to 16 hours after the burst. The image on the right
        includes 9 hours of data starting 3 days and half after
        the burst. The source has faded by a factor 20.
        From~\cite{Costa97}.}
    \label{f:GB0228}
    \end{center}
\end{figure}

 The NFI pointing was performed in 8
hours after the burst time and lasted about 9 hours. In the MECS field
of view a source was found with a flux of $(2.8\pm 0.4)\times 10^{-12}$
erg cm$^{-2}$s$^{-1}$ in the 2-10 keV band \cite{Costa97}. In the ROSAT 
X--ray catalogue, no source was present at that position . The
probability to have by chance a source with that flux in a box of
that size is $\leq 8 \times 10^{-4}$. The field was pointed again three
days after and the source was still there, but had decayed by a factor
20 (see Fig.~\ref{f:GB0228}).

\begin{figure}[!t]
    \begin{center}
    \includegraphics[width=0.6\textwidth,]{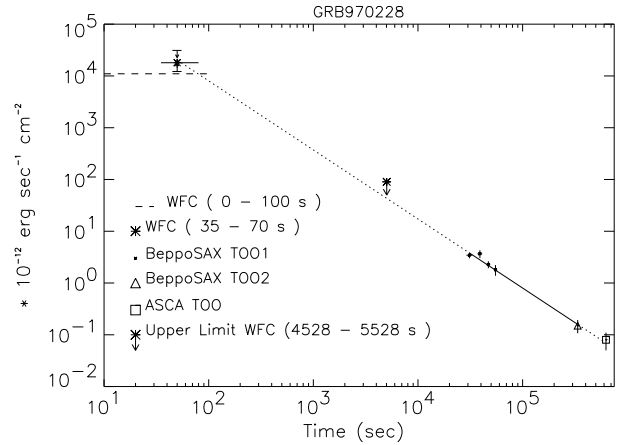}
        \caption{The decay of the GRB970228 from the time of the burst. The
        plot includes the data of the WFC and those of the NFI in the range
        2-10 keV, plus data from ASCA satellite. The decay is described by a power law
        $N(t)\propto t^{-1.33}$. The back extrapolation of this law is
        consistent with the average flux of the prompt burst.
        From Ref.~\cite{Costa97}.}
    \label{f:970228_aft-lc}
    \end{center}
\end{figure}

By subdividing the first observation into three subsets a
light curve was produced, based on NFI data only. 
The source was decaying according to a power law (see Fig.~\ref{f:970228_aft-lc}). 
Quite surprisingly the flux in
the same band detected by the WFC was consistent with the same
power law as well. All the pieces of evidence were there that the fading
X-ray source was the delayed emission, the so called {\em afterglow}, of
the Gamma Ray Burst. Moreover the source position was consistent
with a narrow  confidence strip derived from triangulation of GRBM
and Ulysses data \cite{Hurley97}. The decay of the afterglow source was 
described by a power law $N(t)\propto t^{-1.33}$.

%
%

%
%

While the spectrum of the prompt event
was consistent with the Band function and showed, within each peak,
the already known hard-to-soft evolution, the spectrum of the afterglow
was a power law, $N(E) \propto E^{-2.04}$, constant with
time \cite{Frontera98a}. Both temporal and spectral trends of the
afterglow advocated in favor of a non thermal process and would
be, in the following, the basic building blocks for all the GRB
theories. The afterglow was
studied also by ASCA satellite. We proposed a
pointing of ROSAT, the German X-ray telescope at the end of its
life but, thanks to its high angular resolution, still
the most sensitive X-ray observatory in orbit. The observation was
performed on March 10 and lasted three days. The High Resolution
Imager of ROSAT detected 8 sources but only one was within the
error box of the afterglow of GRB970228 given by
\sax\ \cite{Frontera98b}. With a reasonable hypothesis on the
spectrum to account for different response in energy of the two
instruments, the flux of the ROSAT source was still consistent
with a decay with the same power law. Therefore the association
with the afterglow was very clear and the localization of the GRB
shrunk to ten arcseconds radius.

Meanwhile the coordinates of WFC and those improved of NFIs were
distributed by the BeppoSAX team directly and through circulars to
the International Astronomical Union. Various observers performed
optical observations. Data were reported by five optical
telescopes. The group led by Jan Van Paradijs was the first to
perform two observations of the same field with the same filter.
From the comparison of two images taken with the William Herschel
Telescope on February 28 and other two taken with the same
telescope and with Isaac Newton Telescope on March 8, an optical
point-like source was detected fading from V magnitude 21.3 to V
magnitude $>$ 23.6 \cite{Vanparadijs97}. From an image of the ESO New
Technology Telescope in Chile, the hint of a faint nebulosity was derived. A
subsequent pointing of Hubble Space Telescope showed that the
point-like source had faded down to V magnitude 26.4 in 39
days from the burst and it was embedded in a faint nebular source
of around 25 V mag extended $\sim 1$ arcsec, likely but yet not
necessarily, a galaxy \cite{Sahu97}.

\subsection{The first red-shift}

The further turning point was on May 8, 1997. GRB970508 was
detected by a WFC on GRBM trigger. The follow-up was the fastest
ever performed (5.4 hours after the burst) and the afterglow
was detected by NFIs. An optical counterpart was detected, that,
contrary to GRB970228, had a flux that increased for around two
days, arriving to R mag 20.14 and then started the usual power law
decay. On May 11, when the optical afterglow was still relatively
bright, the CalTech/NRAO group observed it with the Keck Low
Resolution Imaging Spectrograph. Various absorption lines were
identified: some at redshift $z= 0.835$, some other at red-shift of $z
= 0.767$  \cite{Metzger97}. The first of these two values was found
weeks later in emission lines in the extended source that showed
up with the fading of the point source. The mystery of the site of
GRBs was solved. They harbor within remote galaxies.

The immediate consequence was to fix the scale of energetic.  From
the distance we derived the energetics of GRB970508 that, assuming
an isotropic energy emission, resulted to be $E_{iso} =
(0.61\pm 0.13) \times 10^{52}$ ergs.

On December 14, 1997 the redshift of another BeppoSAX GRB (971214)
was determined: $z=3.42$ \cite{Kulkarni98}. The corresponding
energetics was $E_{iso} = (2.45 \pm 0.28) \times 10^{53}$~erg.

But GRB970508 is also relevant for the discovery of the first
radio afterglow with the VLA radio telescope \cite{Frail97}. The
radio emission showed the phenomenon known as
\textit{scintillation} that derives from the effects of
interstellar clouds on sources of very small angular size. In
GRB970508 the scintillation disappeared after around two months.
From the angular size and from the distance, Frail et al. \cite{Frail97} derived the
expansion velocity of the fireball that came out to be around $2c$,
an apparent superluminal effect typical of sources expanding at
relativistic velocity.

\subsection{The Supernova/GRB connection}

Another step forward of the \sax\ era was the first discovery of a
GRB associated  with a SuperNova (SN).  The GRB occurred on 25 April
1998 (GRB\/980425) and was called by us "Liberation Burst". The event was 
not peculiar from the point of view of the Gamma--/X--ray phenomenology, but 
at its location  a supernova (SN1998bw) showed up in the 
optical and radio
band \cite{Galama98b}. The distance was 37 Mpc ($z = 0.0085$) and thence
the energetics was $E_{iso} = (8.5 \pm 0.1) \times 10^{47}$~erg, 
3 to 4 orders of magnitude lower than that of typical
bursts. Beside the positional coincidence, the SN explosion was
(within one day) simultaneous to the GRB and, thence, the latter
was likely the starting event. The SN was classified as Ic and was
unusually bright and characterized by a high
expansion velocity \cite{Patat01}. 

\subsection{The $E_{peak}$ - $E_{iso}$ relation}

Given the remote location of GRBs and the transparency of the Universe 
in X--/gamma--rays, GRBs could be a powerful tool to study those
regions of the remote Universe that cannot be accessed with
optical observation. This possibility was further strengthened
by the discovery of $\emph{dark}$ GRBs, namely of \sax\ 
GRBs for which an optical afterglow was not detected at the level
of other bursts. Since the redshift measurement  is based
on optical/IR spectra, the possibility to use GRBs as beacons
of the obscured Universe relies on the possibility to measure the
intrinsic luminosity of GRBs, on the basis of X or Gamma-ray
measurements only. This is the so called search for a standard
candle. GRBs actually are not standard candles like SNe~Ia and the
energy they release, in the hypothesis of isotropy, spans 3 to 4
orders of magnitude. But if, on the basis of physical and/or
empirical considerations, a link can be found between luminosity
and other parameters measurable in gamma-rays only, the use would
be the same. An important step forward in this direction was the
detection of a strong correlation between the energy $E_{p,i}$ 
at which the intrinsic (i.e., redshift corrected) $\nu F_\nu$ spectrum of the 
GRB peaks and the total energy released during the burst in the hypothesis of 
isotropic emission. This was possible with a first set of \sax\ GRBs followed up with optical
spectrometers \cite{Amati02}. This relationship, now known as 'Amati relation' (from the
name of the first author of our discovery paper \cite{Amati02}), is now confirmed by
subsequent missions (see below).

\section{Theoretical consequences of the \sax\ discoveries. The fireball model}
\label{s:theory}

The cosmological distance scale of GRBs  swept away all the Galactic models. 
The constraints for the extragalactic ones  were a  huge release of  gamma-ray energy 
(up to $\sim 10^{55}$~erg for GRB\,080916C 
assuming isotropy) in a short time (tens of second), non-thermal spectra and short time variability 
(down to ms time scale).
The short time variability $\Delta T$ implies that a huge amount of energy 
is produced in a small volume ($R<c \Delta T$), i.e., the formation of a fireball, a 
concept already introduced in the '70s by Cavallo and Rees \cite{Cavallo78}. However, in a stationary fireball,
due to the high cross section of the interaction $\gamma + \gamma \rightarrow e^+ + e^-$, unlike what is observed, 
the escape of high energy photons ($>1$~MeV) would be suppressed until they have been 
degraded below the pair-production threshold. In addition, due to its high opacity, a stationary fireball 
is expected to emit thermal radiation, while we measure non-thermal spectra. Thus, a relativistically expanding fireball,
already proposed for GRBs in the '80s \cite{Guilbert83,Goodman86,Paczynski86}, became the standard scenario. 
Indeed, if the emitting region is relativistically expanding with Lorentz factor $\Gamma$, the observed photons
are blueshifted by $(1+ \Gamma)$, 
the dimension scale $R$ corresponding to the time variation $\Delta T$ becomes $R<\Gamma^2 c \Delta T$, 
the opacity to electrons associated to baryons and to pairs is decreased by $\Gamma^4$, 
and the threshold for pair production is increased by $\Gamma$ (e.g., Ref.~\cite{Baring97}). Realistic values
of $\Gamma$ that make the optical thickness less than 1 are larger than 100 \cite{Piran99}.
Due to the relativistic beaming, an observer sees the radiation from an angle $\theta = 1/\Gamma$, independently
of the opening angle of the emission.

\begin{figure}[!t]
	\begin{center}
	\includegraphics[width=0.6\textwidth]{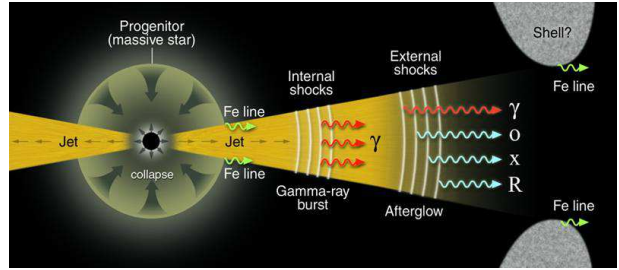}
		\caption{The schematics of the fireball model, also known as standard  model for GRbs. In this figure
a possible progenitor star (a massive star) is assumed. See text.}
	\label{f:fireball}
	\end{center}
\end{figure}

In this scenario, it is expected that, initially, the fireball has a radius $R_{in}$ and energy $E_0$, with a barion 
mass $M_0$  (baryon loading) much lower than $E_0/c^2$. Being the initial optical depth very high,
the radial expansion should be the result of super--Eddington luminosity, in which the internal energy
is converted into bulk kinetic energy. A variant to this scheme is that a fraction of the internal energy is
carried by Poynting flux (e.g., Ref.~\cite{Usov92}). All this energy can be converted into electromagnetic radiation 
through shocks, e.g., between contiguous shells within the fireball (internal shocks) or with the external medium 
(external shocks) (see, e.g., Refs.~\cite{Meszaros94,Paczynski94}). As a consequence, the
bulk Lorentz factor decreases with time. In presence of turbolent magnetic fields expected to be built up behind 
the shocks, the electrons should produce synchrotron radiation \cite{Rees92}. The synchrotron is also expected to
soften as the expanding fireball decelerates and the expected peak, corresponding to the minimum Lorentz factor,
decreases. Thus, while initially the radiation is expected to be emitted in the gamma--ray range (during GRB 
prompt emission), it would progressively evolve into  X--ray afterglow and then UV, optical and radio afterglow. 
A pictorial view of the fireball model is shown in Fig.~\ref{f:fireball}. This model,  
already developed before the \sax\ discovery of the X--ray afterglow, had an immediate success for its capability
to explain the spectral and temporal properties of the GRB afterglows (e.g., Ref.~\cite{Wijers97,Sari98}).  
It became what is now known as GRB standard model. 

As can also be seen from Fig.~\ref{f:fireball}, the fireball model deals with the outer radiating regions, not with
the inner engine, i.e., the source of the relativistic outflow that powers the GRB phenomenon. The issue of the
inner engine was investigated by taking into account the  observed properties.
On the basis of them , in particular the energetics and the variability time scale and GRB duration,
it was suggested that GRBs could be powered by accretion of a massive ($\sim 0.1$ solar masses) accretion disk onto 
a compact object, most likely a black hole, formed as a consequence of the collapse of a 
massive star \cite{Piran99}.

Several progenitors could have given rise to the black hole plus accretion disk: binary 
neutron star or neutron star--black hole mergers \cite{Narayan92}, failed supernovae \cite{Woosley93} or 
collapsar of  hypernovae \cite{Paczynski98}, white dwarf--neutron star mergers.
But also other models were  proposed, like the fireshell model, that links the origin of the energy of gamma-ray
bursts (GRBs) to the extractable energy of electromagnetic black holes \cite{Ruffini01} 
or the supranova model, in which first a supramassive neutron star (SMNS) is formed from the collapse 
of a massive star and, later, for matter accretion or radiative losses of the rotational energy, 
the SMNS collapses to a black hole \cite{Vietri98}. The \sax\ discovery of a redshifted transient Iron K--edge during 
the rise of the 
GRB\,990705 prompt X--ray emission \cite{Amati00} was a strong hint in favor
of the last model and of a model proposed by Berehziani et al. \cite{Berezhiani03}, in which, instead of the late 
formation of a black hole, as  a result of accretion, a transition of the neutron 
star to a deconfined quark star occurs.

%
%

In spite of its success, the standard model leaves some questions
unsolved. 

\begin{figure}[!t]
	\begin{center}
	\includegraphics[width=0.6\textwidth]{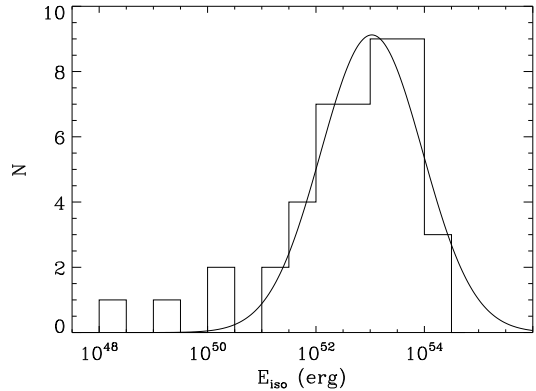}
		\caption{Distribution of the realeased isotropic energy of a sample of GRBs. all the \sax\
events with known redshift are included.
	Reprinted from Ref.~\cite{Amati06}.}
	\label{f:Eiso-distr}
	\end{center}
\end{figure}

One of them  is the energetics. On the basis of the \sax\ 
GRBs with determined redshift, excluding the underluminous GRB980425/SN1998bw, 
the released isotropical energy extended from 
$\sim 5\times 10^{51}$ to $\sim 3\times 10^{54}$~erg. Its distribution, as derived by Ref.~\cite{Amati06}, 
is shown in Fig.~\ref{f:Eiso-distr}. The range of this distribution is now further extended from $10^{50}$ to
$\sim 10^{55}$~erg \cite{Amati09}.
Actually, if the GRB emission is beamed into an angle
$\theta$, the overall energy would be lower by a factor ($1 - \cos{\theta}$).  Indeed Frail et al. \cite{Frail01} and 
Bloom et al. \cite{Bloom03}, attributed the break discovered in the optical afterglow light curves of a number of GRBs
to  beamed afterglow emission, and they derived their beaming angles. Correcting the isotropic
energies for the beaming factor, they found a narrower distribution of the corrected energy $E_\gamma$ with a centroid 
$E_\gamma = 1.3 \times 10^{51}$~erg. The beaming model \cite{Sari99} attributes the break in the light curves as due
when the relativistic beaming angle $\theta_r = 1/\Gamma$ becomes equal to $\theta$, and predicts that the
break in the afterglow light curves would be achromatic. Unfortunately with \sax\, the light curve statistics 
did not allow to test this achromaticity (e.g., Ref.~\cite{intZand01}). 
In spite of the alleviated energy budget if the radiation pattern is beamed, the internal shocks expected in the 
fireball scenario can hardly produce the observed electromagnetic radiation. Indeed, given the similar kinetic energy
of the outflowing shells, only a small fraction ($\sim 10$\%) of this energy can be converted (see, e.g., 
Ref.~\cite{Daigne98}).

Another unsolved question is the energy distribution between GRB prompt emission and its afterglow. The external shocks
(with the external medium) would be more efficient and thus one expects that the energy released in the afterglow 
would be larger than that in the prompt emission. Instead it is found the reverse at least on the basis of the
afterglow measurements now available up to 10 keV. In fact, in the case of GRB\,990123, the only event whose
afterglow was observed with \sax\ PDS up to about 60 keV,  from its high energy spectrum, it was possible to 
estimate a released afterglow energy at least a factor 2 greater than that derived from the 2--10 keV 
spectrum \cite{Maiorano05}.

%

\section{The post \sax\ era}

\subsection{Issues left open by \sax}
\label{s:issues}

In spite of the huge advances obtained with \sax,  many questions about
GRBs could be answered only with further X--/gamma--ray observations, as also discussed by 
one of us  in 2003 \cite{Frontera03}. The most relevant open questions were: 
\begin{itemize}
\item
{\bf X--ray afterglow light curve at early times and breaks at late times}
 
The first question could not be solved by \sax\ due to the unavoidable time ($>\sim$ 5 hrs)
needed to start follow--up X--ray observations of GRBs with the NFIs.
The break detection in the X--ray afterglow light curves required more sensitivity.
The problem was: are the breaks observed in the optical band  achromatic? 
In order to test the jet model, it was of key importance to measure X-ray afterglow light curves with higher
statistical quality.

\item
{\bf Afterglow of short GRBs}

No short GRB was detected  with \sax\ WFCs and thus no follow-up of 
these events to search for possible afterglow sources was performed.

\item
{\bf GRB environment} 

If the GRBs are the result of the collapse of a massive star, the GRB environment
could be Fe polluted and
line features could be  likely found. Actually, with \sax\ we found evidence of a significantly variable $N_H$ 
in the prompt emission \cite{Frontera04a} and, as above discussed, of a clear redshifted Fe 
absorption K--edge \cite{Amati00}. 

In addition, it was found evidence of Fe X--ray emission lines at 3--4 $\sigma$ level during the afterglow phase
of a few GRBs \cite{Piro99,Antonelli00,Yoshida99,Piro00}.
All these results required a confirmation with more sensitive telescopes.  

\item
{\bf Dark GRBs}

While most of the well localized GRBs (about 90\%) showed  X--ray afterlow \cite{Frontera03}, only  30\% of them
had an optical counterpart \cite{Fynbo01}, even if there was an indication of a larger fraction from a later
analysis \cite{Depasquale03}. The questions were: does the optical emission
have not been observed because of a fast decay? Or are these GRBs intrinsically weak? Or is their brightness
optically obscured by presence of dust in the environment around them or are they not optically visible 
due to a high redshift of the object? Only fast afterglow searches, starting soon
after the main event with a multiple wavelength coverage of the afterglow
emission, could  settle this issue.

\item
{\bf Origin of X--ray flashes} 

A number of \sax\ GRBs  did not have a gamma--ray counterpart \cite{Heise01} and they were called
X--Ray Flashes (XRF). Were these events 
at very high redshifts (z$>$10) or GRBs with intrinsically soft spectra? 
Also in this case, more sensitive observations of these events were requested.

\item 
{\bf Emission mechanisms at work during the prompt and afterglow emission}

The time averaged spectra of many GRBs
were consistent with an optically thin synchrotron shock model 
\cite{Tavani96,Amati01}. However there was a small number of GRBs for which
this model did not work. While for an optically thin synchrotron, the expected power-law index 
of the $E F(E)$ spectrum below 
the peak energy $E_p$ cannot be steeper than 4/3 (ideal case of an instantaneous
spectrum in which the electron cooling is not taken into account, see, e.g., \cite{Frontera00a}
and references therein), in many cases the measured spectra, even those time resolved, 
are inconsistent with these expectations. To overcome these difficulties either modifications 
of the synchrotron scenario \cite{Lloyd00} or other radiative models were suggested: e.g., a synchrotron 
self-Compton model \cite{Meszaros00,Stern04}, Compton up-scattering 
of low energy photons by a quasistatic plasma \cite{Liang97}, superposition of 
blackbody spectra \cite{Blinnikov99}, a Compton drag emission model \cite{Lazzati00}. 
Each of these models was capable to interpret some of  the spectral properties, but failed to 
interpret others. 

Also the X--ray afterglow spectra seemed to be consistent with a synchrotron 
origin \cite{Galama98a}. However in some cases multiwavelength spectra (from radio to gamma) 
showed that the X--ray component could be due to Inverse Compton of synchrotron radiation emitted 
at lower frequency \cite{Harrison01}.

\end{itemize}

\subsection{X--/gamma--ray missions of the post \sax\ era}

The HETE 2 mission was launched during the \sax\ life, on 9 October 2000, but was fully 
operational from 2002 \cite{Ricker03}. One of its most important results was the accurate
localization of a short GRB, allowing for the first time the detection of a GRB optical afterglow
and redshift \cite{Villasenor05}.

The intriguing questions left open by \sax\ motivated also the approval of new missions
devoted to GRBs. The first one 
is \swift, a NASA mission that was launched on November 2004 and 
fully operational in April 2005 \cite{Gehrels05a}. 
Another NASA mission, devoted to study, among other goals, GRBs, is GLAST
({\em Gamma-Ray Large Area Space Telescope}) launched on 11 June 2008, and
 after the launch, renamed \fermi\ \cite{Atwood09}.
All these missions have international collaborations, with relevant contributions from 
Italy. 
Other missions, like the European 
{\it International Gamma-Ray Astrophysics Laboratory} ({\it INTEGRAL}) \cite{Winkler03} 
and the Italian {\it Astro-rivelatore Gamma a Immagini LEggero} 
({\it AGILE})\cite{Basset07} are also giving contributions to GRB
studies.
Apart from HETE 2, all these missions are still operational.

\subsection{The most relevant results obtained after  \sax}

The mission that has given the highest contribution to the GRB afterglow study 
and to the GRB redshift measurement is \swift, that performs immediate (within $\sim$ 1 min) 
X-ray and optical follow-up observations of GRBs detected with a Burst Alert Telescope (BAT) aboard
the same satellite. The alert triggers are also immediately distributed
for prompt optical, near-infrared, and radio observations with ground-based
telescopes. 
\swift\ was designed mainly to study the early afterglow by filling the gap left by \sax\ between
the prompt emission and the late afteglow ($> 5$ hrs from the GRB onset). The most important results
are summarized below.

\subsubsection{The early afterglow}

\begin{figure}[!t]
	\begin{center}
	\includegraphics[width=0.6\textwidth]{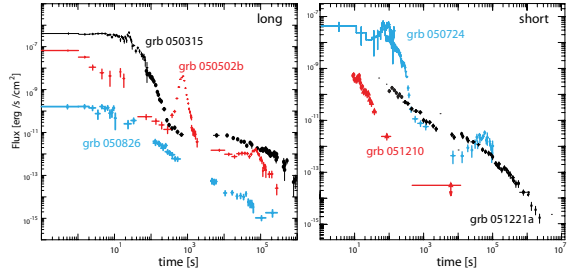}
		\caption{Examples of representatives X-ray light curves obtained with \swift. {\em Left
	panel}: long GRBs; {\em right panel}: short GRBs. Reprinted from Ref.~\cite{Gehrels09}.}
	\label{f:swift_lc}
	\end{center}
\end{figure}

One the most striking results of \swift\ has been the determination of the early X--ray (0.5-10 keV)
afterglow light curve. Examples of representative afterglow light curves obtained with \swift\
are shown in Fig.~\ref{f:swift_lc}.

%
%

As can be seen, long GRBs at early times show two main types of afterglow light curves: steep to shallow 
light curves, which are the most frequent, and smoothly declining light curves. In the first case 
X--ray flares are often observed (in about 30\% of the cases; \cite{Chincarini10}). The first X-ray flare
in the afterglow was found with \sax\ from an X--ray flash (XRF011030) \cite{Galli06}.

\begin{figure}[!t]
	\begin{center}
	\includegraphics[angle=-90,width=0.6\textwidth]{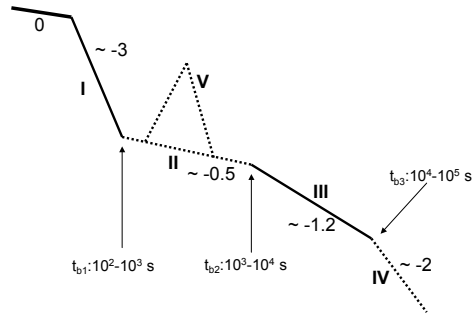}
		\caption{Canonical afteglow light curve since early times derived from \swift\ data. 
	Reprinted from Ref.~\cite{Zhang06}.}
	\label{f:canonical-lc}
	\end{center}
\end{figure}

From these results a canonical shape of the GRB afteglow light curves was proposed \cite{Nousek06}, now well
confirmed \cite{Gehrels09}. This canonical shape is shown in Fig.~\ref{f:canonical-lc}, where the phase ''0'' 
denotes the prompt emission, while the afterglow is denoted by 4 segments: the first and the third 
are the most common, while the other three(dashed lines) are observed only in a fraction of bursts. 
The typical temporal indices in the four segments are shown  in the figure. The photon  spectrum does not change 
from segment II to IV (see Fig.~\ref{f:canonical-lc}) : it is a power-law ($I(E)\propto E^{-\alpha}$) 
with a typical value of $\alpha$ of 2, coincident
with the mean value found from the \sax\ late afterglow data \cite{Frontera03},  while the
spectrum of  Segment I is softer and in general similar to the flare spectra \cite{Zhang06}.
At late times ($10^4$--$10^5$~s), a further time break is observed. This break could coincide with that observed
in the optical band during the \sax\ era, on the basis of which the beaming factor was derived. Instead, for \swift\ 
GRBs contemporarily observed in the X--ray and optical bands, it is found that these breaks are in general not 
simultaneous and thus are not achromatic, as expected by the beaming model \cite{Campana07}.

The most accepted interpretation of the early canonical light curve is that the initial steep 
segment is the tail of the prompt emission \cite{Kumar06}, the  steep-to-shallow transition occurs when the 
afterglow radiation, due to an external forward shock, becomes dominant over the prompt emission. The long flatness 
of the shallow segment could be the result of  a continuous energy injection \cite{Nousek06,Zhang06}.  
The second break is interpreted as a result of the exhaustion of the energy injection. The fact that the spectra 
of both flares and steep segment are similar is a strong hint that the mechanism that gives rise to flares is 
the same the gives rise to the prompt emission. 

In order to interpret the chromaticity of the breaks, Ghisellini et al. \cite{Ghisellini07} propose that, 
after the steeply decaying
phase that follows the early prompt, the electromagnetic radiation is the sum of two emission components: 
the "late-prompt" emission (due to late internal shocks), and the "real afterglow" emission (due to external shocks).
In this scenario,  the X-ray emission from the onset of the shallow decay phase and onwards 
could be  dominated by "late prompt" emission,  due to prolonged activity of the central source, while the optical
emission during the same time is dominated by the external forward shock emission (i.e. the traditional afterglow
emission).

The efficiency crisis  of the fireball shock model discussed above is confirmed by the \swift\ results:
the energy in the afterglow is lower or, at most, comparable to  that in the prompt emission, except that we are
underestimating the afterglow energetics (see the warning discussed in Sect.~\ref{s:theory}).

%
%

\subsubsection{Localization and afterglow of short bursts}

With \swift\ it has been eventually possible to accurately localize short bursts and to detect 
their afterglow emission (e.g., Ref.~\cite{Gehrels05b}). The afterglow properties can fall within the general scheme 
discussed above for long GRBs (see two representatives examples in Fig.~\ref{f:swift_lc}). 
Also short GRBs show flares during the afterglow phase. The flares typically happen hundreds
of seconds after the trigger or earlier, but in some cases they occur around a day after the trigger 
(e.g., GRB 050724 \cite{Barthelmy05}, see Fig.~\ref{f:swift_lc}).
Also the properties of the optical afterglows are similar to those of long GRBs. In general short bursts have
similar peak fluxes of long GRBs and higher peak energy $E_p$. 

Due to the still low number of short GRBs with known $z$ (about a dozen, \cite{Amati10}), their redshift distribution
is still an open issue. Initially it seemed that short bursts were located in the local Universe ($z<1$).
However the discovery of a redshift of $z =6.7$ for GRB 080913 \cite{Greiner09}, a burst with a $T_{90}$ duration 
of 8~s in the observer frame, but an intrinsically short burst ($<2$~s), makes
the assumption of a local Universe for short bursts untenable.
 
\begin{figure}[!t]
	\begin{center}
	\includegraphics[width=0.6\textwidth]{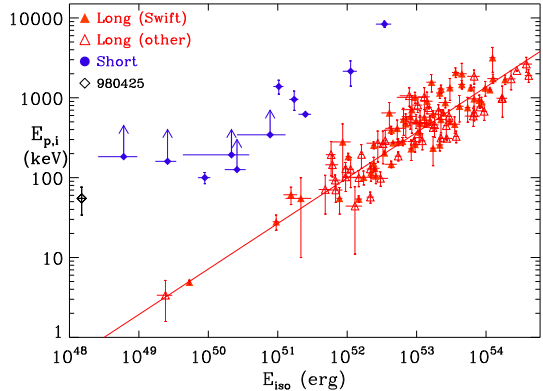}
		\caption{The $E_{p,i}$--$E_{iso}$ relation for long (Amati relation) and
	short bursts. Reprinted from Ref.~\cite{Amati10}.}
	\label{f:amati-short_long}
	\end{center}
\end{figure}

An important result concerns the $E_{p,i}$--$E_{iso}$ correlation. It is found \cite{Ghirlanda09,Amati10} that
short GRBs do not follow the Amati relation derived for long bursts, but, in the $E_{p,i}$--$E_{iso}$ plane, 
occupy a region parallel to that occupied by the long GRBs (see Fig.~\ref{f:amati-short_long}).
 
Another important \hete\ result, soon after confirmed with \swift, was the apparent detection of an extended emission 
after  short events \cite{Villasenor05}, confirming the results found with BATSE (e.g., Ref.~\cite{Lazzati01}) and with
\sax\ GRBM \cite{Montanari05}.

%
%

\subsubsection{Circumburst medium}

The determination of the circumburst medium is crucial also to establish the progenitor star.
The evidence of emission lines during the afterglow phase found during the \sax\ era has not been confirmed
by \swift\ \cite{Hurkett08}. 
Instead, as far as  Fe K--edge and variable absorption during the prompt emission, after \sax\ no other
still operational satellite 
devoted to study the GRB prompt emission has covered the X--ray band  down to 2 keV. This gap
should be covered with future missions.

From the low energy cut-off of the \swift\ afterglow X--ray spectra, it has been  possible to  determine the GRB
$N_H$, and, with ground optical telescopes, in many cases it has been
possible to detect  damped Lyman-alpha lines and  metallic absorption lines of the same GRB or its host (see, e.g., 
Refs.~\cite{Sokolov01,Savaglio06}). It is found that
the neutral hydrogen column density so measured  does not agree with the one inferred from the X-ray absorption. 
This type of discrepancy may be ascribed either to systematic errors in the spectra interpretation \cite{Butler07} or
it is real, and thus phenomena, like ionization and/or a higher metallicity, could be a possible explanation. This
issue is still matter of investigation.

\subsubsection{Dark GRBs}

Thanks to the \swift\ accurate (arcsec) and prompt (within minutes) localization of GRBs 
(about 500 events already detected), it has been possible to establish how many of them are optically dark.
The result is that the fraction of non detected afterglows is still 30--40\% \cite{Greiner11}.
From the comparison of the spectra measured in the X--ray band and in the near-infrared/optical band
it appears that the faint optical emission could be the combination of two effects: a moderate intrinsic 
extinction for GRBs with low redshift, and a distance effect for GRBs
at high redshifts ($>$5). The latter events are estimated  to be about 20\% of dark GRBs \cite{Greiner11}.

\subsubsection{Origin of X--ray flashes (XRFs)}

On the basis of the \sax\ and \hete\ results, there are strong pieces of evidence that X--ray flashes are low luminosity GRBs. 
Their properties are consistent with GRBs with lower $E_p$ and higher fluxes in X--rays than in gamma--rays. 
The other spectral parameters of the prompt emission and the temporal behaviour of their afterglows at various 
wavelengths are identical to those of  GRBs \cite{Dalessio06}. Also an analysis of X--ray flashes in 
the complete \hete\ spectral catalog \cite{Pelangeon08} shows that most XRFs lay at low/moderate redshift 
and that they are likely predominant in the GRB population and closely linked to GRBs.
A confirmation of their link with GRBs is their consistency with the Amati relation \cite{Amati07}.

\subsubsection{The Amati relation after \sax}

Thanks to the many flying missions, we have a much larger sample of GRBs with well determined
redshift. For those events for which has been possible to derive, along $z$, their bolometric fluence 
and energy peak $E_p$ (about 100), it has been possible to estimate both the intrinsic 
(redshift corrected) $E_{p,i}$ and the isotropic energy $E_{iso}$.  

As shown in Fig.~\ref{f:amati-short_long}, all these GRBs are fully consistent with the Amati relation, inclusive of 
the most energetic events recently discovered with \fermi\ \cite{Amati09}, even though some authors, 
due to its spread, suspect that
this relation could be influenced by selection effects (see, e.g., Ref.~\cite{Butler09}). However other 
authors (e.g., Ref.~\cite{Ghirlanda08}) find that these effects do not invalidate the relation. Also the time 
resolved spectra, obtained by slicing the GRB time profile in several time intervals and deriving 
the spectra in each of them, show a correlation between time resolved  $E_p$ and corresponding 
flux \cite{Frontera10}.

The physics underlying this relation has been investigated by various authors, but it is still matter of debate.
Some authors \cite{Zhang02,Panaitescu09} have attempted to give an interpretation of this relation
in the context of the fireball shock model. Thompson et al. \cite{Thompson07} propose 
that this relation could be the result of blackbody radiation, but the test of this model \cite{Ghirlanda07}
has given negative results: the blackbody, if present, is only of modest relative intensity 
with respect to the non thermal component.

\subsubsection{GRB/supernova connection}

Prompt optical, infrared and radio observations of \hete\ first and thus  \swift\ GRBs have definitely confirmed that
several long GRBs orginate in supernova explosions. Nowadays we know a dozen of supernovae clearly associated to GRBs, 
but there is evidence of several optical counterparts of GRBs that show late rebrightenings, 
interpreted as SNe emerging from the afterglow light curves \cite{DellaValle06b}. In general these SNe are type Ibc with
high expansion velocities and much larger energy release  than  in  normal SNe (these peculiar SNe are also 
dubbed {\em hypernovae}, HNe). 
However there are GRBs not associated with SNe (see, e.g., Ref.~\cite{DellaValle06a}), demonstrating that 
there are GRBs originating in very faint supernovae or they are 
due to different phenomena. But there are also cases of HNe with any GRB associated
with them \cite{Guetta07}. In these cases, the most likely interpretation of this case is that the GRB emission 
is beamed along the rotation axis of the progenitor star and that, depending on the angle between the line of sight 
and the GRB emission direction, we can see both events (SN explosion and GRB) or only one of the them.

\subsubsection{High energy gamma--rays ($>100$ MeV) from GRBs}

Observations of GRBs at high energies ($>$100 MeV) are quite important. They can solve important issues like
to distinguish between hadronic and leptonic origin of the gamma-ray radiation, to probe for signatures 
within GRBs of the acceleration processes of the Ultra High Energy Cosmic Rays, or to establish the 
bulk Lorentz factor of the fireball (see, e.g., Ref~\cite{Band09}).

Mainly with  \fermi\ and \agile\ satellites, high energy gamma--rays ($>$100 MeV) have been observed from  GRBs and
their mean properties studied (see, e.g., Ref.~\cite{Zhang11}).
Merging together high energy ($>100$~MeV) and low energy data ($>8$ keV), the time resolved spectra  
are best modeled with the Band function over the entire \fermi\ spectral range, 
which may suggest a common origin of the low energy and high energy emissions. In some cases, however, 
the superposition of a blackbody component plus an extra power--law components is the best description.
The origin of these components are still matter of debate. 
An intriguing result is the discovery of a time lag of a few seconds in the arrival of the high energy emission
with respect to that at low energies, also matter of debate.

\section{Open issues and perspectives}

The prompt localization of GRBs and the consequent afterglow discovery  has opened a powerful window of 
investigation of the Universe. The potentiality of the \sax\ discovery has been demonstrated by the impressive 
results already obtained with \swift\ first and now with \fermi\ and \agile. \swift\ has allowed 
to study the early afterglow. In addition it has allowed 
to discover the afterglow of the short events and to establish their host galaxies. A great result of the
 \swift\ era is the discovery of the most distant events, up to $z\sim 9.4$ (GRB\,090429, \cite{Cucchiara11}). 
This has shown the potentialilty of GRBs for cosmology studies.
In spite of these huge advances, specially in the study of the GRB afterglow, the GRB phenomenon is still 
far to be understood. It is now recognized that only going back to the study of the 
prompt emission is possible to understand the original explosion. 

Also the models developed thus far to interpret the prompt and/or the afterglow emission
are not capable to explain all the observed properties. An example is the difficulty of interpreting the
Amati relation. A confirmation of the theoretical difficulties is the title of a recent paper 
by Maxim Lyutikov \cite{Lyutikov09}: "Gamma Ray Bursts: back to the blackboard".

\subsection{Some relevant open issues}

Some of the most relevant still open issues are the following:
\begin{itemize}
\item {\bf Physics underlying the GRB prompt emission}

While the basic properties of the afterglow emission (fading law of the flux and 
power-law spectral shape) can be satisfactorily explained in the framework of the "standard" fireball 
plus external shock scenario, the physics underlying the complex light curves and the fast spectral 
evolution of the "prompt" emission is still far to be understood.  

Presently, there is a "forest" of models invoking different kinds of fireball (kinetic energy dominated, 
Poynting flux driven), of shocks (internal, external) and of emission mechanisms (synchrotron and/or Inverse 
Compton originated in the shocks, direct or Comptonized thermal emission from the fireball photosphere, 
and mixtures of these).

\item {\bf Circumburst environment and GRB progenitors}

 The nature of the GRB progenitors  is still an open issue. 
The classical scenario is that short GRBs are connected with the merging of compact objects, while
long GRBs are most probably connected to the collapse of massive fast rotating stars 
(e.g., Ref~\cite{Paczynski98}).

In spite that this scenario continues to be the most successful, we do not have sufficient
information to freeze it.
For example, the collapsar model, in which a black hole forms due to the 
initial failing of the supernova explosion and the GRB originates from a jet emerging along 
the rotation axes, predicts that the revitalized Supernova 
explodes simultaneously to the GRB. This prediction is still not accutately tested, and
alternative models, like the supranova model \cite{Vietri98} or the quark star or  hybrid hadron-quark star 
model \cite{Berezhiani03} or, still, the new model proposed by Titarchuk et al. \cite{Titarchuk11}, in which a very 
massive star ($>$100 solar masses) explodes with no formation of a compact object, cannot be excluded.
The study of absorption K--edges of heavy elements and/or variable absorption could provide 
strong help  to identify the real scenario.

For that, very broad band observations of the prompt emission, from a fraction of keV to GeV are crucial. 
Also the detection of non-electromagnetic radiation (high energy neutrinos, gravitational radiation) associated
to the prompt gamma-ray event could be of key importance to identify models and establish the central engine.

\item {\bf GRB polarization}

It is recognized the crucial importance of polaziation measurements from GRBs to test their
emission models. The low value of polarization detected in the optical afterglow suggests that 
even the most sensitive future X--ray experiments could not detect any polarization if the
mechanism at work, e.g., synchrotron, is the same in both bands. 
The situation is completely different for the prompt emission, where  high polarization levels
have been reported, but one of these results is controversial
\cite{Coburn03}, while  others (evidence of a transient polarization; e.g., \cite{Gotz09}) require
a confirmation. Various satellite/ISS/balloon experiments devoted to this issue are in progress.

\item {\bf Cosmology with GRBs}

If, on one hand, Cosmic Microwave Background (CMB) allows to observe the Universe at $z= 1000$ and its
primordial fluctuations in the matter density at that time, 
we still do not know when these density fluctuations have produced gravitationally-bound systems,
whose internal evolution has given rise to dark matter, visible Universe (stars, galaxies, quasars)
and its reionization.
The formation of the first  objects should have taken place at epochs corresponding 
to $z= $10-30, certainly beyond $z= 7$.  GRBs are thus the most potential sources for exploring this extreme 
Universe. In less than 15 years, they have permitted to observe the Universe  up to $z= 9.4$, 
while in 50 years, the most distant quasar discovered is at $z = 6.96$. 

Attempts to use GRBs as cosmological rulers have already been performed by Amati et al. \cite{Amati08} exploiting
the Amati relation, and  by Ghirlanda et al. \cite{Ghirlanda04b} exploiting the $E_{p,i}$--$E_\gamma$ relation,
which is derived from the Amati relation by correcting for the beaming angle. 
\end{itemize}

\subsection{Perspectives}

The GRB field is still very young, so many other exciting discoveries will be certainly done in the next future.
In addition to the facilities already available at multiwavelength, like \swift, \fermi, \agile\ and
the ground optical and radio telescopes, for GRB studies other missions are being 
studied, or are under development or just operative. They include space X--/gamma--ray missions, 
Cerenkov facilities and new radio facilities.

Among the space X--/gamma--ray missions, it merits to be mentioned the French--Chinese mission SVOM \cite{Schanne10} for
the study of the prompt emission in the 4 keV--5 MeV energy band. 
Missing pieces of the GRB jig-saw could be provided 
by hard X--/soft gamma--ray telescopes based on multi-layers and Laue optics that could study the 
afterglow at energies above 10 keV, by wide field polarimeters to search for polarization of 
the prompt emission, and focal plane polarimeters to search for polarization of the afterglow.  

Current Cerenkov facilities include  MAGIC, HESS, and VERITAS devoted  to detect high 
energy ($\sim 1$~TeV) gamma--ray photons. They may play a key role especially with their band extension to 
lower energies where the major transparency for pair production on interstellar background light 
extends the horizon to far away objects. The future facility CTA will also benefit of a
larger FOV.   

Among the radio facilities it merits to be mentioned the Expanded VLA (EVLA) to 
study the radio counterparts of GRBs down to 10~$\mu$Jy flux density.
Other possibilities are offered by wide or very  wide FOV observatories such as LOFAR and SKA.

Also non-electromagnetic facilities, like those devoted to detect neutrinos (AMANDA, IceCube, ANTARES) or gravitational waves 
(LIGO, TAMA, VIRGO) include among their goals GRBs, even if they appear still not so sensitive to detect signals
from GRBs at cosmological distances. 

In conclusion, the discovery of the GRB sites and afterglow with \sax\ will continue to push up the frontiers
of our knowledge of the Universe for many years.

\acknowledgments
Many people contributed, also in a fundamental way, to the \sax\ success for the GRB discoveries. 
We wish to thank all of them. The \sax\ mission was a joint program of the Italian 
space agency ASI and of the Netherlands Agency for Aerospace Programs.

\bibliography{bibl_NC}   
\bibliographystyle{varenna}

\end{document}